\documentclass[12pt]{article}
\newcommand{\beq}{\begin{equation}}
\newcommand{\eeq}{\end{equation}}
\newcommand{\bea}{\begin{eqnarray}}
\newcommand{\eea}{\end{eqnarray}}

\newcommand{\gsim}{\lower.7ex\hbox{$
\;\stackrel{\textstyle>}{\sim}\;$}}
\newcommand{\lsim}{\lower.7ex\hbox{$
\;\stackrel{\textstyle<}{\sim}\;$}}

\newcommand{\eod}{\end{document}}

\begin{document}
\thispagestyle{empty}
\vspace*{-22mm}

\begin{flushright}
UND-HEP-12-BIG\hspace*{.08em}05\\
%Version 3.0 \\
%\today\\

%hep-ph/0703132\\

\end{flushright}
\vspace*{1.3mm}

\begin{center}
{\Large {\bf Probing CP Violation in $\tau ^- \to \nu (K\pi/K2\pi / 3K/ K3\pi )^- $ Decays}}

\vspace*{10mm}

{ I.I.~Bigi$^a$}\\
\vspace{7mm}
$^a$  {\sl Department of Physics, University of Notre Dame du Lac, Notre Dame, IN 46556, USA}\\

{\sl email addresses: ibigi@nd.edu} \\

\vspace*{10mm}

{\bf Abstract}
\vspace*{-1.5mm}
\\

\end{center}

It had been suggested to probe CP violation in $\tau \to \nu K\pi $ decays with 
$K^0 - \bar K^0$ oscillation to produce $A_{\rm CP}(\tau^- \to \nu K_S\pi^- ) = (0.36 \pm 0.01)\%$. 
BaBar has found $A_{\rm CP}(\tau^- \to \nu K_S\pi^- [\geq \pi^0]) = (-0.36 \pm 0.23 \pm 0.11)\%$ -- 
i.e., 2.8 sigma difference with SM prediction. It is discussed, why one needs to probe 
$A_{\rm CP}(\tau \to \nu K\pi)$, $A_{\rm CP}(\tau \to \nu K2\pi)$ and $A_{\rm CP}(\tau \to \nu K3\pi )$ 
{\em separately} to establish the `existence' of New Dynamics and its `features'.  
It should be possible at SuperB \& Super-Belle experiments.

\vspace{3mm}

\hrule

\tableofcontents
\vspace{5mm}

\hrule\vspace{5mm}

%%%%%%%%%%%%%%
\section{CP Violation in Leptonic Dynamics}
%%%%%%%%%%%%%%%%%

Large CP asymmetries have been established in $B_d$ transitions by BaBar, Belle and CDF experiments, and the first evidence has appeared for $A_{\rm CP}(D^0 \to K^+K^-) - A_{\rm CP}(D^0 \to \pi^+ \pi^-)$ \cite{LHCBD,CDFD}. While we have found that CKM 
dynamics give at least the leading source of CP violation in $B$ transitions, it cannot contribute significantly to the  
"matter" vs. "anti-matter" asymmetry in `our' universe.  

New Dynamics (ND) have been found by neutrino oscillations with 
$\theta_{12}, \theta_{23}, \theta_{13} > 0$ \cite{DAYA,RENO}. A {\em necessary} condition for 
generating CP violation there has been satisfied.  It gives a good chance for leptonic dynamics producing `matter' vs. `anti-matter' asymmetry as a `shadow' effect of the `lepton' vs. `anti-lepton' asymmetry. 
Furthermore probing CP symmetry at the level of ${\cal O}(0.1 \%)$ in $\tau \to \nu [K\pi/K2\pi]$ has roughly the 
same sensitivity of ND in the amplitude as searching for BR$(\tau \to \mu \gamma)$ at the level of $10^{-8}$ 
\cite{CPBOOK,72}. For CP odd observables in a SM allowed decay are {\em linear} in a ND amplitude, while 
in SM forbidden ones the rates are {\em quadratic} in ND amplitudes: 
\beq
{\rm CP \; odd} \propto T^*_{\rm SM} T_{\rm ND}  \; \; \; \;  vs. \; \; \; \; {\rm LFV} \propto |T_{\rm ND} |^2
\eeq
Leptonic EDMs and CP asymmetries in $\mu$ decays have been probed with high sensitivities 
\cite{72} and should be continued. 

Now BaBar Collaboration has found some evidence for CP violation in $\tau$ decays \cite{BABARTAU}: 
\beq 
A_{\rm CP}(\tau^- \to \nu K_S\pi^- [\geq \pi^0] ) = (-0.36 \pm 0.23 \pm 0.11)\% 
\eeq 
CP violation established in $K^0 - \bar K^0$ oscillations gives as predicted \cite{BSTAU,NIRTAU} 
(whether it is given by CKM dynamics or not): 
\beq 
A_{\rm CP}(\tau^- \to \nu K_S\pi^- )= (0.36 \pm 0.01)\% \; ;   
\eeq 
i.e., there is a difference of 2.8 sigma between these two values. 

There is some experimental sign of {\em global} CP violation in $\tau$ decays. 
However, global asymmetries are often much reduced. One needs to probe different final states -- include 
three- and four-body ones to established its (or their)  
{\em existence} of ND. Furthermore it is crucial to determine its (or their) {\em features}. 
We should  
focus on transitions that are CKM suppressed in SM -- like $\tau \to \nu [K+ \pi's] $ -- where one has a good 
chance to identify both the impact and features of ND with less `background' from SM amplitudes. 

One needs conceptual lessons to understand the basis of the observed data on 
$\tau^- \to \nu K_S\pi^- [\geq \pi^0]$ vs.  $\tau^- \to \nu [K\pi ]^- $, $\tau^- \to \nu [K2\pi]^- $,  
$\tau^- \to \nu [3K]^- $ and $\tau^- \to \nu [K3\pi]^- $: 
\begin{itemize}
\item 
CP asymmetries in $\tau \to \nu [K_S+ \pi's] $ are generated by measured $K^0 - \bar K^0$ oscillations with great accuracy. 
One can measure rates and CP violations in $\tau ^- \to \nu [K\pi's]^- $ vs. 
$\tau ^+ \to \bar \nu [K +\pi's]^+$ 
and to calibrate ratios of  $\tau ^- \to \nu [\pi's]^- $ vs. $\tau ^+ \to \bar \nu [\pi's]^+$, where one expects that 
ND can hardly produce measurable asymmetries. 
\item 
For $\tau ^- \to \nu [K\pi]^-$ one gets contributions mostly from $\tau^- \to \nu K^*(892)$ with some from 
$\tau^- \to \nu K_0^*(1430)$ due to vector and scalar exchanges.  

\item 
For $\tau \to \nu [K2\pi]/ \nu [3K] /\nu [K3\pi] $ one has more CP odd observables through {\em moments} and their {\em distributions} to check the impact of ND. Those are described by 
total four- \& five-body  final states -- and therefore {\em hadronic} three- \& four-body  final states 
with {\em distributions} of hadronic masses \cite{CPBOOK,MIRANDA,2MIRANDA,MIKE}. In particular 
one should probe $K^*(892)\pi$, $K_1(1270)$, $K_1(1400)$ \& 
$K^*(1410)$ hadronic final states and in particular their {\em interferences}. 

\item 
Beyond $K^0 - \bar K^0$ oscillations one probes {\em direct} CP violation in $\tau$ decays. Unless one has 
{\em longitudinally polarized} $\tau$, one needs differences in both the weak and strong phases to 
generate CP asymmetries in $\tau \to \nu [K\pi]$. Non-zero T odd observables can be produced by 
FSI without CP violation. On the other hand true CP asymmetries can be probed for $\tau^-$ vs. $\tau ^+$ decays. 

\item
CPT symmetry predicts 
\beq
\Gamma (\tau ^- \to \nu +[S=-1]) = \Gamma (\tau ^+ \to \bar \nu +[S= 1])
\eeq
with 
\bea
[S=-1] &=& \bar K^0 \pi^-/K^- \pi^0/\bar K^0 \pi^-\pi^0/K^-\pi^+\pi^-/
K^-\pi^0\pi^0/\\
&& K^-K^+K^-/K^-\bar K^0 K^0/ \bar K^0(3\pi)^-/K^-(3\pi)^0 \; \; {\rm etc.}
\eea
Two items have to be dealt with:
\begin{itemize}
\item
One measures final states with $K_S$, $K_L$ and the interferences between them. $K^0 - \bar K^0$ 
oscillation impacts CP asymmetries as expressed by 2Re$\, {\epsilon_K }$ in a {\em global} way for channels. 
\item
Mixing between $\bar K^0 \pi^- \to K^- \pi^0$, $\bar K^0 \pi^0 \to K^- \pi^+$ and  
$K^- K^+ \to K^0 \bar K^0$ happen by FSI. Diagrams show it, but we cannot control it quantitatively.

\end{itemize}

\end{itemize}
Therefore one can learn crucial lessons about the underlying dynamics by  identifying those final states 
{\em separately}. The branching ratios of these transitions are not small \cite{PDG}.

%%%%%%%%%%%%%%
\section{CP Asymmetry in $\tau ^- \to \nu [K\pi]^- $}
%%%%%%%%%%%%%%%%

One has three-body final states with two hadrons $h_{1}$ \& $h_{2}$ with variation in 
$M^2(h_1h_2)$ with vector and scalar resonances. CP asymmetries depend on different weak and strong 
phases.

Final states $K\pi$ are produced from the QCD vacuum with vector and scalar configurations with 
form factors $F_V$ and $F_S$ \cite{PICH}; the vector component  is dominated mainly in the form of $K^*$. 
In principle the latter produces no problem, since several resonances contribute at different 
mass values. In the SM one gets no different weak phase from quark and lepton dynamics -- however 
ND can generate different weak phases due exchanges of charged Higgs or the `old standby' for 
enhanced ND effects, namely SUSY with broken $R$ parity. Amplitudes with scalar resonances are 
suppressed. Therefore their contributions are hardly to be found for total widths; however they can 
generate {\em interference} with vector resonances with `local' CP asymmetries up to of ${\cal O}(\% )$ 
\cite{DEL}. 
Therefore one has to probe the `topology' of the three-body final states for 
$\tau^- \to \nu K_S\pi^- $ vs. $\tau^+ \to \bar \nu K_S\pi^+ $ \cite{CLEO} and 
$\tau^- \to \nu K^-\pi^0 $ vs. $\tau^+ \to \bar \nu K^+\pi^0 $ by $d\Gamma/dE_K$ or 
$d\Gamma/dM_{K\pi}$ etc.  

The `Miranda procedure' has been suggested for three-body final states for $B$ and $D$ decays 
for {\em localizing}  
CP asymmetries in Dalitz plots \cite{MIRANDA,2MIRANDA}. It can be applied here independent of $\tau$ 
production asymmetry with plots of  $E_K$ vs. $M_{K\pi}$. In particular 
one can compare regions with positive and negative interference between vector and scalar states for 
$\tau^-$ vs. $\tau ^+$, which gives significant lessons on the underlying ND. `Miranda procedure' is based 
on analyzing the {\em significance}  
\beq
\Sigma (i) \equiv \frac{N(i) - \bar N(i)}{\sqrt{N(i) + \bar N(i)}}
\eeq
in the final state plot rather the customary {\em fractional} asymmetry
\beq
\Delta (i) \equiv \frac{N(i) - \bar N(i)}{N(i) + \bar N(i)} \; .
\eeq 

At a SuperB experiment proposed and approved near Rome in Italy one could produce a 
pair of longitudinally polarized $\tau$ and therefore probe T {\em odd} moments and their distributions 
in $\tau \to \nu h_1h_2$ decays.

%%%%%%%%%%
\section{CP Violation in $\tau \to \nu h_1h_2h_3/\nu h_1h_2h_3h_4$}
%%%%%%%%%

Final states with three or four hadrons in the final state produce many more CP sensitive observables. 
Therefore we have more information about the existence and the features of the ND and check also experimental uncertainties  \cite{CPBOOK,72}. 
In the SM one gets zero CP asymmetries in $\tau ^- \to \nu K^- [S=0]^0$ and only a global 
one in $\tau ^- \to \nu K_S [S=0]^-$ due to 2Re$(\epsilon_K)$. ND in those decays has to compete only with SM Cabibbo suppressed ones.

%%%%%%%%%%%%%%
\subsection{CP Asymmetry in $\tau ^- \to \nu [K2\pi]^-  /\nu [3K]^- $}
%%%%%%%%%%%%%%%  

For $A_{\rm CP}(\tau ^- \to \nu K^- \pi^+\pi^- )$ one predicts global zero CP asymmetry in SM and 
$(0.36 \pm 0.01)\%$ as before due to $K^0 - \bar K^0$ oscillation for $A_{\rm CP}(\tau ^- \to \nu K_S\pi^- \pi^0)$. 
A richer landscape for ND can surface in $\tau \to \nu  K2\pi $ due to contributions from  
$K^*\pi$, $K\sigma$, $\kappa \pi$ etc., where one sees triple-product asymmetries \cite{DATTA}.  
To be more practical: One can measure {\em T odd moments}  
$\langle \vec p_K \cdot (\vec p_{\pi_1} \times \vec p_{\pi_2}  )  \rangle$ for $\tau^-$ vs. $\tau ^+$ decays.  

One can also probe their Dalitz plots with one refinement: the total mass of the 
hadronic final state is not fixed -- it depends on the energy of the neutrino. 
One can follow the qualitative example given in $K_L \to \pi^+\pi^-e^+e^-$ transitions 
\cite{SEHGAL} (and suggested for $D^0 \to K^+K^-\pi^+\pi^-$ \cite{CPBOOK}). Final states with 
three hadrons produce huge fields for CP observables even for unpolarized $\tau$ leptons: their 
Dalitz plots can be probed depending on the energy of the neutrino. This can be seen as an 
`excess of riches'. However one has to think which observables give us the `best' lessons about the 
underlying dynamics in the `real' world. For example, one can focus on measuring the angle between 
the plane of the two hadrons and the plane of the neutrino and the third hadron in the $\tau$ rest frame  -- like $\pi^+-\pi^-$ and $\nu - K^-$:    
\bea
\frac{d}{d\Phi_{+-}}\Gamma(\tau ^- \to \nu K^-\pi^+\pi^-) &=& \Gamma^{K^-}_1 {\cos}^2\Phi_{+-} + 
\Gamma^{K^-}_2{\sin}^2\Phi_{+-} + \Gamma^{K^-}_3{\cos}\Phi_{+-} {\sin}\Phi_{+-}\\
 \frac{d}{d\Phi_{-+}} \Gamma(\tau^+\to \bar \nu K^+\pi^-\pi^+ )&=& \bar{\Gamma}^{K^+}_1
 {\cos}^2\Phi_{-+} + \bar{\Gamma}^{K^+}_2{\sin}^2\Phi_{-+} +  \bar{\Gamma}^{K^+}_3 {\cos}\Phi_{-+} {\sin}\Phi_{-+} 
%\frac{d}{d\Phi_K}\Gamma(\tau ^- \to \nu K_S\pi^-\pi^0) &=& \Gamma^{K_S}_1\; {\cos}^2\Phi_K + 
%\Gamma^{K_S}_2\;{\sin}^2\Phi_K + \Gamma^{K_S}_3\; {\cos}\Phi_K \;{\sin}\Phi_K \\
% \frac{d}{d\Phi_K} \Gamma(\tau^+\to \bar \nu K_S \pi^+\pi^0 )&=& \bar{\Gamma}^{K_S}_1\; {\cos}^2\Phi_K + 
% \bar{\Gamma}^{K_S}_2\;{\sin}^2\Phi_K -  \bar{\Gamma}^{K_S}_3\; {\cos}\Phi_K \;{\sin}\Phi_K \; .
\eea
Using $\Phi_{+-} = - \Phi_{-+}$, cos$^2 \Phi_{+-}=$ cos$^2 \Phi_{-+}$, 
sin$^2 \Phi_{+-}= $sin$^2 \Phi_{-+}$,  cos$\Phi_{+-}$sin$\Phi_{+-}$ = - cos$\Phi_{-+}$sin$\Phi_{-+}$ 
one gets: 
\bea
\Gamma (\tau ^- \to \nu K^-\pi^+\pi^-) - \Gamma(\tau^+\to \bar \nu K^+\pi^-\pi^+ ) &=& 
\frac{\pi}{2} \left([\Gamma_1 - \bar \Gamma_1] + [\Gamma_2 - \bar \Gamma_2]  \right)
\\
\int_0^{\pi/2}d\Phi_{+-} \left( \Gamma_{\tau^-} + \Gamma_{\tau^+}  \right) - 
\int_{\pi/2}^{\pi}d\Phi_{+-} \left( \Gamma_{\tau^-} + \Gamma_{\tau^+}  \right) &=& 
\Gamma_3 + \bar \Gamma_3 \; ;  
\eea
i.e., `global' CP asymmetry 
$\Gamma (\tau ^- \to \nu K^-\pi^+\pi^-) \neq \Gamma(\tau^+\to \bar \nu K^+\pi^-\pi^+ )$ and the first step 
towards `local' CP violation. 

The next step for localizing CP asymmetry is to measure 
\beq
\Gamma_1 \neq \bar \Gamma_1 \; , \; \Gamma_2 \neq \bar \Gamma_2 \; , \; 
\Gamma_3 \neq \bar \Gamma_3 \; . 
\eeq 
Strong QCD forces can generate $0\neq \Gamma_3 = - \bar \Gamma _3$; however CP violation 
shows $\Gamma_3 + \bar \Gamma_3 \neq 0$. Measuring $\Gamma_{1,2,3}$ \& $\bar \Gamma_{1,2,3}$ 
{\em separately} with cos$^2\Phi_{+-}$, sin$^2\Phi_{+-}$ and cos$\Phi_{+-}$ sin$\Phi_{+-}$ also help experimental uncertainties. 

Furthermore one can measure the angles between the planes of $K^- -\pi^+$ and $\nu - \pi^-$ vs. 
$K^+ -\pi^-$ and $\bar \nu - \pi^+$ or $K^- -\pi^-$ and $\nu - \pi^+$ vs. 
$K^+ -\pi^+$ and $\bar \nu - \pi^-$ in $\tau^-$ and $\tau^+$ decays. Those angles 
$\Phi_{K^-\pi^+}$ vs. $\Phi_{K^+\pi^-}$ or $\Phi_{K^-\pi^-}$ vs. $\Phi_{K^+\pi^+}$ tell us more 
of the underlying dynamics in $\tau \to \nu K\pi\pi$ transitions. 

Likewise one can probe $\tau^- \to \nu K^-K^+K^-$ vs. $\tau^+ \to \bar \nu K^+K^-K^+$. It is more 
challenges for $\tau^- \to \nu K^- \pi^0\pi^0$ vs. $\tau^+ \to \bar \nu K^+ \pi^0\pi^0$ and 
$\tau^- \to \nu K_S \pi^-\pi^0$ vs. $\tau^+ \to \bar \nu K_S \pi^+\pi^0$.

As the final step of probe CP asymmetries one can measure the {\em distributions}  
of $\vec p_K \cdot (\vec p_{\pi_1} \times \vec p_{\pi_2}  )$ with $d\Gamma/dM_{K\pi}$ and 
$d\Gamma/dM_{K2\pi}$ or with $d\Gamma/dM_{2K}$ and/or $d\Gamma/dM_{3K}$. 

As mentioned before the sizable number of kinematical variables and of specific channels 
allow more internal crosschecks of systematic uncertainties like detection efficiencies for positive vs. 
negative particles. In addition: 
\begin{itemize}
\item 
ND can interfere {\em vector} with {\em axial vector} configurations. For example, ND could be based on 
$W_R$ exchanges coupling to right-handed leptons and quarks. 
\item 
Quantitative correlations of $\Gamma_3^K - \bar \Gamma_3^K$ with $\Gamma_3^{\pi} - \bar \Gamma_3^{\pi}$ are 
important. These should be possible at SuperB/Super-Belle.     
\end{itemize} 

`Miranda Procedure' can be applied as mentioned above; it is driven by data for `partitioning' the Dalitz plots.  
However one needs some refinement: The $K2\pi$ and 
$3K$ masses are not fixed -- they depend on the neutrino kinematics impact. Dividing the hadronic 
mass spectrum into two or three parts could help significantly depending on future data. 

%%%%%%%%%%%%%%
\subsection{CP Asymmetry in $\tau ^- \to \nu [K3\pi]^- $}
%%%%%%%%%%%%%%%

The landscape is even richer for impact of ND for these final states from 
$K^*\rho$, $K^*\sigma$, $K\omega$, $\kappa \rho$ etc. 
There are one several 
different T odd moments for 
$\tau ^- \to \nu K^- \pi^+\pi^-\pi^0$ 
\beq
\langle \vec p_{K^-} \cdot (\vec p_+ \times \vec p_- )   \rangle \; \; , 
\; \; \langle \vec p_{K^-} \cdot (\vec p_{\pi^+} \times \vec p_{\pi_0} )   \rangle \; \; {\rm etc.}
\eeq 
and for $\tau ^- \to \nu K_S \pi^-\pi^+\pi^-$
\beq
\langle \vec p_{K_S} \cdot (\vec p_{\pi^+} \times \vec p_{\pi^-} )   \rangle \; \; , 
\; \; \langle \vec p_{K_S} \cdot (\vec p_{\pi^-} \times \vec p_{\pi^-} )   \rangle \; \; {\rm etc.}
\eeq
and for distributions in $M_{h_1h_2}$, $M_{h_1h_2h_3}$ and $M_{h_1h_2h_3h_4}$. Of course it will 
need more experimental work -- but it would tell us more of the features of ND.

%%%%%%%%%%%%
\section{Summary}
%%%%%%%%%%%%

SM cannot generate measurable CP asymmetries in $\tau ^- \to \nu [K^- + \pi's]$ and  
a value of $(0.36 \pm 0.01) \%$ in total widths for $\tau ^- \to \nu [K_S + \pi's]$. ND -- like 
with charged Higgs or $W_R$ exchanges -- can affect 
these decays with hadronic two-, three- and four-body final states significantly with probing 
regions of interference between different resonances. To be more precise: 
\begin{itemize}
\item 
One has to compare $\Gamma (\tau^- \to \nu [K\pi ]^-)$ vs. $\Gamma (\tau^+ \to \bar \nu [K\pi ]^+)$, 
$\Gamma (\tau^- \to \nu [K 2\pi ]^-)$ vs. $\Gamma (\tau^+ \to \bar \nu [K2\pi ]^+)$, 
$\Gamma (\tau^- \to \nu [3K ]^-)$ vs. $\Gamma (\tau^+ \to \bar \nu [3K ]^+)$ and 
$\Gamma (\tau^- \to \nu [K3\pi ]^-)$ vs. $\Gamma (\tau^+ \to \bar \nu [K3\pi ]^+)$. 

\item 
As emphasized before about $B$ and $D$ decays with three- and four-body final states, one 
gets contributions from resonances and their interferences for CP asymmetries. 
However `global' asymmetries averaged over the total widths are significantly smaller than 
individual contributions.  
\item 
Therefore it is very important to probe the `topology' in the Dalitz plots. 

\item 
For $\tau^- \to \nu [K\pi ]^-$ one can probe interference between vector and scalar states, 
which are somewhat suppressed. 
For $\tau^- \to \nu [K2\pi ]^-/[3K]^-$ one can probe T odd moments due to vector and axial vectors 
exchanges and even more for $\tau^- \to \nu [K3\pi ]^-$, which should be not suppressed in general. 

\item 
On the step to probe the final states as discussed above one can look for {\em local} asymmetries in 
$\tau ^- \to \nu [3K + K2\pi ]^-$ vs. $\tau ^+ \to \bar \nu [3K + K2\pi ]^+$.

\end{itemize}

SuperB and Super-Belle experiments should be able to probe the {\em whole} area of 
$\tau \to \nu [K\pi/K2\pi/3K/K3\pi]$ transitions with {\em neutral} pions in the final states. 

One more comment about CP asymmetries in $\tau$ decays: These comments about the  
impact of ND is focused on 
semi-hadronic $\tau$ transitions. It is most likely to affect also $B$ and $D$ decays -- but it 
could `hide' more easily there due to larger effects -- in particular for $B$ transitions -- and less control over 
non-perturbative QCD effects.  

A last comment: I have emphasized to probe the distributions of final states in $\tau$ (and $B/D$) decays to 
find the impacts and the features of ND(s) based on `binned' \cite{MIRANDA,2MIRANDA} and 
`unbinned multivariate' \cite{MIKE} results.

\vspace{0.5cm}

{\bf Acknowledgments:} This work was supported by the NSF under the grant number PHY-0807959. 

\vspace{4mm}

%%%%%%%%%%%%%%%%

\end{document}